\documentclass[runningheads,a4paper]{llncs}

\usepackage{amssymb}
\usepackage{graphicx}

\usepackage{url}
\newcommand{\keywords}[1]{\par\addvspace\baselineskip
\noindent\keywordname\enspace\ignorespaces#1}

\begin{document}

\mainmatter  

\title{Optimal scales in weighted networks}


%
%
\author{Diego Garlaschelli $^{1}$ 
\and Sebastian E. Ahnert $^{2}$
\and Thomas M. A. Fink $^{3}$
\and Guido Caldarelli $^{4,3,5}$}
\authorrunning{Diego Garlaschelli et al.}

\institute{$^{1}$ Lorentz Institute of Theoretical Physics, University of Leiden,\\ Niels Bohrweg 2, 2333 CA Leiden, The Netherlands\\
$^{2}$ Cavendish Laboratory, University of Cambridge,\\ 
JJ Thomson Avenue, CB3 0HE Cambridge, United Kingdom\\
$^{3}$ London Institute for Mathematical Sciences,\\ 22 South Audley St, W1K 2NY London, United Kingdom\\
$^{4}$ IMT Alti Studi Lucca, Piazza S. Ponziano 6, 55100 Lucca, Italy\\
$^{5}$ ISC-CNR, Dipartimento di Fisica, Universit\`a La Sapienza,\\ P.le A. Moro 2, 00185 Roma, Italy
}

%
%

\toctitle{Optimal scales in weighted networks}
\tocauthor{Diego Garlaschelli, Sebastian E. Ahnert, Thomas M. A. Fink, Guido Caldarelli}
\maketitle

\begin{abstract}
The analysis of networks characterized by links with heterogeneous intensity or weight suffers from two long-standing problems of arbitrariness.
On one hand, the definitions of topological properties introduced for binary graphs can be generalized in non-unique ways to weighted networks.
On the other hand, even when a definition is given, there is no natural choice of the (optimal) scale of link intensities (e.g. the money unit in economic networks).
Here we show that these two seemingly independent problems can be regarded as intimately related, and propose a common solution to both. 
Using a formalism that we recently proposed in order to map a weighted network to an ensemble of binary graphs, we introduce an information-theoretic approach leading to the least biased generalization of binary properties to weighted networks, and at the same time fixing the optimal scale of link intensities. We illustrate our method on various social and economic networks.
\end{abstract}

\section{Introduction}
A large number of social, economic, biological and information systems can be conveniently described as networks (or graphs) where $N$ nodes (or vertices) are connected by $L$ links (or edges).
Over the last fifteen years, Network Science has emerged as a fast-growing discipline crossing the boundaries of many research fields \cite{guidosbook}.
The aim of Network Science is that of characterizing and modelling the structure and dynamics of real-world networks, as opposed to abstract mathematical specifications such as those studied by Graph Theory.

One of the challenges in Network Science is that of extending the relatively well-developed tools available for \emph{binary networks} (where links are either present or absent, with no possible variation in their intensity) to the more general case of \emph{weigthed networks} (where links can have heterogeneous weights) \cite{vespy_weighted,kertesz_clustering,newman_weighted,fagiolo_clustering,myensemble}.
For instance, in binary social networks a link may represent the existence of a friendship relation between two people, irrespective of the strength of such a relation, while in weighted social networks a link may be attached a value indicating the amount of shared time or the degree of intimacy between two friends.
While the analysis of social networks has traditionally focused on binary graphs, the recent availability of detailed data about the magnitude of  interactions in large-scale social systems offers a new potential for the study of social networks as weighted graphs.
However, in the transition from binary graphs to weighted networks two main problems of arbitrariness are encountered, and are still largely unsolved.

First, while several definitions of basic topological quantities have been introduced for binary graphs, the corresponding generalizations to weighted networks are non-unique. 
An important example is that of the \emph{clustering coefficient} $c_i$, defined in binary undirected graphs as the fraction of neighbours of node $i$ that are also neighbours of each other, or equivalently the fraction of triangles in which node $i$ participates \cite{guidosbook}.
In weighted networks, the clustering coefficient can be generalized in many ways \cite{vespy_weighted,kertesz_clustering,newman_weighted,fagiolo_clustering,myensemble}, and there is no natural criterion indicating the optimal definition.
Another example is the \emph{reciprocity}, defined in binary directed graphs as the ratio of reciprocated to total links \cite{myreciprocity}.
In weighted networks, there are many possible generalizations requiring sophisticated comparisons and calculations \cite{mywreciprocity}.
In general, on one hand the heterogeneity of link intensity observed in weighted networks provides important additional information that one would like to exploit in order to define generalized quantities that reduce to the ordinary and well-studied ones in the particular case of binary graphs, but on the other hand a large degree of arbitrariness makes the problem not well defined.

Second, even when a definition of a weighted quantity is given, one is left with the problem of the arbitrary scale of link intensities. 
The simplest example is perhaps the \emph{total weight} $W$ of a network, defined as the sum of all link weights in the graph.
If the links of the network represent e.g. flows of money between the units of an economic system, or the time spent by two friends in their phone calls, the quantity $W$ clearly depends on the units chosen (e.g. Euros or thousands of Euros, minutes or seconds, etc.).
Similarly, any other quantity depending on the edge weights will suffer from the same arbitrariness.
This problem can be circumvented by defining adimensional weights that are invariant under rescaling, e.g. dividing each edge weight by the average weight over all pairs of vertices.
However, this still does not solve the problem entirely.
Consider for instance, as one of the simplest properties of binary graphs,  the \emph{link density} defined as the ratio of the number of observed links to the total number of pairs of vertices.
This quantity ranges between zero (empty graph) and one (fully connected network).
The corresponding weighted quantity, if defined as the ratio of the total weight $W$ to the number of pairs of vertices, ranges between zero and infinity and thus loses the properties of a density.
This problem persists irrespective of the preliminary rescaling of the edge weights. 
Similar considerations apply to the \emph{global} clustering coefficient defined as the fraction of realized triangles: in weighted networks, the weighted counterpart of such a `fraction' can actually range from zero to infinity.

In this paper, we show that the two seemingly unrelated problems discussed above can actually be rephrased as two sides of the same coin.
In sec. \ref{sec:ensemble} we first briefly recall a general method that we proposed in order to generalize the definition of any topological property valid for binary graphs to one valid for weighted networks \cite{myensemble}.
While powerful, this approach still does not uniquely fix the scale of edge weights and the functional form of the mapping from binary to weighted properties.
For these reasons, in sec. \ref{sec:statphys} we show that this approach can be rephrased within a statistical physics formalism fixing the functional form of the mapping \cite{mytemperature}.
Then, in sec. \ref{sec:maxent} we apply the Maximum Entropy principle to further fix the scale of edge weights 
in such a way that the weighted topological properties induced by the binary ones are defined in the least biased way.
As a result, we obtain an information-theoretic method that fixes the optimal scale of edge weights in the original network and at the same time induces unique and least biased definitions of weighted properties from the well-known binary ones.
In sec.\ref{sec:real} we finally illustrate our method on various real-world social and economic networks.

\section{Weighted Networks as Ensembles of Binary Graphs\label{sec:ensemble}}
Mathematically, a binary directed network with $N$ vertices is uniquely specified by a $N\times N$ \emph{adjacency matrix} $\mathbf{A}$ with entries $a_{ij}=1$ if a directed link from vertex $i$ to vertex $j$ is present, and $a_{ij}=0$ otherwise. 
For binary undirected networks, where links have no orientation, the matrix $\mathbf{A}$ is symmetric.
Weighted directed networks are instead characterized by a $N\times N$ \emph{weight matrix} $\mathbf{W}$ where the (non-negative, for the purposes of this article) entry $w_{ij}$ represents the intensity of the directed link connecting vertex $i$ to vertex $j$ (including $w_{ij}=0$ if the link is absent).
Again, in weighted undirected networks 
the matrix $\mathbf{W}$ is symmetric.
In this paper, we will consider directed networks, where it is intended that  undirected networks can be obtained as the special situation where $\mathbf{A}$, $\mathbf{W}$ and other similar quantities are symmetric.

Quite recently \cite{myensemble}, we proposed a method to extend any definition of topological property valid for binary graphs, i.e. any function $f^{(b)}(\mathbf{A})$ of the binary adjacency matrix $\mathbf{A}$, to a corresponding function $f^{(w)}(\mathbf{W})$ of the weight matrix $\mathbf{W}$. 
Our method is based on the idea that the matrix $\mathbf{W}$ specifying the original weighted network can be mapped to an ensemble of binary graphs defined by a conditional probability $P(\mathbf{A}|\mathbf{W})$. 
The latter represents the occurrence probability, given $\mathbf{W}$, of a possible graph $\mathbf{A}$ in the ensemble.
This mapping from $\mathbf{W}$ to $P(\mathbf{A}|\mathbf{W})$ allows one to define the weighted counterpart $f^{(w)}(\mathbf{W})$ of any binary property $f^{(b)}(\mathbf{A})$ as the expected value of the latter over the ensemble of binary graphs, i.e.
\begin{equation}
f^{(w)}(\mathbf{W})\equiv \langle f^{(b)}(\mathbf{A})\rangle_\mathbf{W}=\sum_\mathbf{A} P(\mathbf{A}|\mathbf{W}) f^{(b)}(\mathbf{A})\: .
\label{eq:mapping}
\end{equation}

If we require that each edge weight $w_{ij}$ only determines the probability $p_{ij}=p(w_{ij})$ of existence of a binary link from vertex $i$ to vertex $j$ (while having no effect on a different pair of vertices), then $P(\mathbf{A}|\mathbf{W})$ simply factorizes over pairs of vertices, i.e.
\begin{equation}
P(\mathbf{A}|\mathbf{W})
=\prod_{i,j}[p(w_{ij})]^{a_{ij}}[1-p(w_{ij})]^{1-a_{ij}}
\label{eq:factor}
\end{equation}
where $i<j$ for undirected networks and $i\ne j$ for directed networks with no self-loops (if self-loops are allowed, then we should set $i\le j$ for undirected networks and no constraint for directed networks).
The problem then reduces to specifying the functional form of the (monotonic) edge-specific probabilities $p_{ij}=p(w_{ij})$ \cite{myensemble}.
If these probabilities are regarded as entries of a $N\times N$ matrix $\mathbf{P}(\mathbf{W})$, the factorized form (\ref{eq:factor}) allows to considerably simplify the definition of any weighted properties given in eq.(\ref{eq:mapping}).
For instance, for any quantity $f^{(b)}(\mathbf{A})$ that is polynomial or multilinear in the entries $a_{ij}$ of the adjacency matrix, the corresponding weighted property reduces to \cite{myensemble}
\begin{equation}
f^{(w)}(\mathbf{W})\equiv \langle f^{(b)}(\mathbf{A})\rangle_\mathbf{W}= f^{(b)}[\mathbf{P}(\mathbf{W})]\: .
\label{eq:mapping2}
\end{equation}

In our first approaches to the problem \cite{myensemble,myensemble2}, we chose the linear mapping
\begin{equation}
p(w_{ij})\equiv \frac{w_{ij}-w_{min}}{w_{max}-w_{min}}
\label{eq:linear}
\end{equation}
where $w_{min}$ and $w_{max}$ represent the minimum and maximum observed weight in the network, respectively.
The above choice ensures that $p(w_{ij})$, as required in order to be a probability, ranges between $0$ and $1$.
We showed that this approach can effectively exploit the additional topological information encoded
in the weights, in particular for fully connected networks \cite{myensemble,myensemble2}.
However, eq.(\ref{eq:linear}) violates two desirable properties of $p(w_{ij})$, namely $p(0)=0$ and $p(+\infty)=1$, i.e. the fact that (only) missing links in the original network are associated with zero connection probability in the binary ensemble, and that (only) infinite connection intensities in the original network are associated with unit connection probability.

In general, the choice of the functional form of $p(w_{ij})$ remains somewhat arbitrary, and eq.(\ref{eq:linear}) can be viewed as the mathematically simplest possibility.
This translates the arbitrariness of the initial problem, i.e. the non-uniqueness of the generalization of a binary topological property to a weighted counterpart, to the arbitrariness of the choice of $p(w_{ij})$.
This also implies that the second problem of arbitrariness, i.e. the fact that any weighted topological property $f^{(w)}(\mathbf{W})$ has in general an undesired dependence on the choice of the units of $\mathbf{W}$ in the orginal network, is still unsolved.
While the linear choice in eq.(\ref{eq:linear}) is invariant under changes of units (i.e. it is scale-invariant), this will not be the case for more general non-linear choices of $p(w_{ij})$.

\section{Statistical Physics of Network Ensembles\label{sec:statphys}}
We now show that the above approach can be rephrased within a statistical physics formalism in such a way that the first arbitrariness, i.e. the choice of the functional form of $p(w_{ij})$, can be fixed.

Very recently \cite{mytemperature}, we introduced a general ensemble of binary graphs that, as in statistical physics, is defined (here in slightly simplified form) by the occurrence probability
\begin{equation}
P(\mathbf{A})=\frac{1}{\mathcal{Z}}\exp\left[\frac{-E(\mathbf{A})}{T}\right]\: .
\label{P_general_t}
\end{equation}
In the above equation,  $E(\mathbf{A})$ is the \emph{energy} of the particular graph $\mathbf{A}$ (a function of one or more topological properties of $\mathbf{A}$, representing the `cost' of realizing that graph), $T$ is the \emph{temperature} (representing the degree of topological optimization, with lower $T$ corresponding to a probability concentrated on energetically `cheaper' configurations) and
\begin{equation}
\mathcal{Z}\equiv\sum_{\mathbf{A}}\exp\left[\frac{-E(\mathbf{A})}{T}\right]
\end{equation} 
is the normalizing constant, or \emph{grand partition function} of the ensemble.
Graph ensembles like the one defined above are extensively used in the statistical physics literature \cite{mywreciprocity,mytemperature,newman_expo,mybosefermi,ginestra}
as well as in social science \cite{wasserman,snijders}, where they are known as $p^*$ models or \emph{Exponential Random Graphs}.

Since $E(\mathbf{A})$ represents the cost of realizing the particular graph $\mathbf{A}$, we can regard the ensemble of binary graphs discussed in sec. \ref{sec:ensemble} and defined by the probability $P(\mathbf{A}|\mathbf{W})$ as a particular case of the ensemble defined by eq.(\ref{P_general_t}) where the energy is a function $E(\mathbf{A},\mathbf{W})$ of the weight matrix $\mathbf{W}$ \cite{mytemperature}.
In particular, the requirements for $P(\mathbf{A}|\mathbf{W})$ leading to the factorized form (\ref{eq:factor}) translate into the requirement of the additivity of $E(\mathbf{A},\mathbf{W})$, i.e.
\begin{equation}
E(\mathbf{A},\mathbf{W}) \equiv \sum_{i,j} \epsilon_{ij}a_{ij}= \sum_{i,j} \epsilon(w_{ij})a_{ij}
\label{eq:local}
\end{equation} 
where, again, $i\ne j$ for directed networks and $i<j$ for undirected networks. 
In the above expression, $\epsilon_{ij}=\epsilon(w_{ij})$ must be interpreted as an edge-specific energy, i.e. the energetic cost contributed by the existence of a link from vertex $i$ to vertex $j$ ($a_{ij}=1$).
In this way, the choice of the functional form of $p(w_{ij})$ translates to the choice of the functional form of $\epsilon(w_{ij})$.
Indeed, it is easy to show that inserting eq.(\ref{eq:local}) into eq.(\ref{P_general_t}) leads precisely to eq.(\ref{eq:factor}) where
\begin{equation}
p(w_{ij})=\frac {e^{-\epsilon(w_{ij})/T}}{1+e^{-\epsilon(w_{ij})/T}}
\label{eq:p}\: .
\end{equation}

The above expression is particularly useful in order to select the appropriate form of $\epsilon(w_{ij})$. Specifically, we see that a linear dependence of the type $\epsilon(w_{ij})\propto w_{ij}$ is not suitable, since it would assign a probability $p(0)=1/2$ (rather than $p(0)=0$) to the pairs of vertices connected by no link ($w_{ij}=0$) in the original weighted network.
We also see that the linear choice (\ref{eq:linear}) is not natural, since it would correspond to a very complicated, and difficult to justify, form of $\epsilon(w_{ij})$.
On the other hand, as we recently noted \cite{mytemperature}, the simplest satisfactory choice involves a proportionality between $e^{-\epsilon(w_{ij})/T}$ and $w_{ij}$, i.e. $e^{-\epsilon(w_{ij})/T}=z w_{ij}$ or in other words
\begin{equation}
p(w_{ij},z)\equiv\frac{zw_{ij}}{1+zw_{ij}}\: .
\label{eq:fermi}
\end{equation}
This means that the dependence of the binary link energy on the observed edge weight is given by
\begin{equation}
\epsilon(w_{ij},z)=-T\ln(z w_{ij})\: ,
\label{eq:log}
\end{equation}
i.e. $w_{ij}$ has a logarithmic effect on $\epsilon(w_{ij},z)$.
In real networks with a power-law weight distribution of the form $\rho(w)\propto w^{-\alpha}$, the above relation can be used to measure the empirical temperature as $T=\alpha-1$ \cite{mytemperature}.
Typical observed values are $0.5\lesssim T\lesssim 2.5$.

Equation (\ref{eq:fermi}) fixes the functional form of $p(w_{ij},z)$ in a very reasonable manner.
With such a choice, we recover, for all values of $z$, the desired properties $p(0,z)=0$ and $p(+\infty,z)=1$.
Note that if $z= [w_{max}-w_{min}]^{-1}$ and $zw_{ij}\ll 1$ then we have $p(w_{ij},z)\approx w_{ij}/[w_{max}-w_{min}]$, which is approximately equivalent to the choice in eq.(\ref{eq:linear}). 
This corresponds to a `sparse graph' limit for the binary ensemble induced by the weighted network.
However, in general the value of $z$ in eq.(\ref{eq:fermi}) is arbitrary. This leads us to the main point of this paper, which is discussed in the next section.

\section{Maximum-Entropy Scale of Edge Weights\label{sec:maxent}}
We can regard the arbitrariness of $z$ in eq.(\ref{eq:fermi}) as equivalent to the arbitrariness of the unit of edge weights in the original network.
Indeed, changing the scale of $w_{ij}$ to $\lambda w_{ij}$, where $\lambda$ is any positive constant, is mathematically equivalent to changing $z$ to $\lambda z$.
In particular, from eqs.(\ref{eq:fermi}) and (\ref{eq:log}) it is clear that
\begin{equation}
p(\lambda w_{ij},z)=p(w_{ij},\lambda z)\quad\textrm{and}\quad \epsilon(\lambda w_{ij},z)=\epsilon(w_{ij},\lambda z)\: .
\end{equation}
This shows that the scale $\lambda$ can be completely reabsorbed in a redefinition of the parameter $z$, i.e. $z\to \lambda z$.
Therefore, without loss of generality, we can regard $z$ in eq.(\ref{eq:fermi}) as the parameter specifying the scale of weights. 
If we introduce a unique way to fix $z$, we have automatically eliminated the second and last source of arbitrariness discussed in the Introduction, i.e. the units of edge weights in the original network.

In what follows, we propose the Maximum Entropy principle as a rigorous criterion to fix the value of $z$, and further show that this value is unique.
Our main idea is that, in line with other uses of the Maximum Entropy principle \cite{newman_expo,jaynes}, the least biased choice of a quantity should correspond, in absence of any other indication, to the one that maximizes Shannon's entropy (given the available information).
Given a real-world weighted network specified by the matrix $\mathbf{W}$ and the corresponding binary ensemble specified by the conditional probability $P(\mathbf{A}|\mathbf{W})$ as given by eqs.(\ref{eq:factor}) and (\ref{eq:fermi}), Shannon's entropy reads
\begin{equation}
S(z)\equiv-K\sum_\mathbf{A} P(\mathbf{A}|\mathbf{W})\ln P(\mathbf{A}|\mathbf{W})
\label{SP}
\end{equation}
where $K$ is an arbitrary constant, that we fix later for convenience. 
Now, due to the factorization of $P(\mathbf{A}|\mathbf{W})$ as in eq.(\ref{eq:factor}), and since the entropy of a factorized process is additive, we can simply write
\begin{equation}
S(z)=K\sum_{i,j}s_{ij}(z)
\label{eq:ssum}
\end{equation}
(with the usual convention on $i,j$ for directed and undirected graphs) where $s_{ij}(z)$ is the edge-specific entropy 
\begin{equation}
s_{ij}(z)=-p(w_{ij},z)\ln p(w_{ij},z)-[1-p(w_{ij},z)]\ln[1-p(w_{ij},z)]
\label{eq:slocal}
\end{equation}
Note that both missing links ($w_{ij}=0$) and very large weights ($w_{ij}\to+\infty$) generate a zero entropy $s_{ij}(z)=0$, and therefore have no effect on the choice of the optimal scale. 
This is consistent with the fact that both zero and infinite weights are independent of any chosen scale $\lambda$.
Inserting eq.(\ref{eq:slocal}) into eq.(\ref{eq:ssum}), we find that the entropy of the ensemble is
\begin{equation}
S(z)=-K\ln\prod_{i,j} [p(w_{ij},z)]^{p(w_{ij},z)}[1-p(w_{ij},z)]^{1-p(w_{ij},z)}
\label{eq:sfinal}
\end{equation}
If we want $S(z)$ to be normalized between $0$ and $1$ (although this has no effect on the following results), we can set
\begin{equation}
K\equiv \frac{1}{M\ln 2}
\end{equation}
where $M$ is the number of possible pairs of vertices, i.e. $M=N(N-1)$ for a directed network with no self-loops and $M=N(N-1)/2$ for an undirected network with no self-loops. If self-loops are allowed, then the above values of $M$ must be increased by $N$.

We can now look for the value of $z$ that maximizes $S(z)$ as given by eq.(\ref{eq:sfinal}).  
To this end, we write the first derivative of $S(z)$ as
\begin{eqnarray}
S'(z)&=&
K\sum_{i,j}\frac{\partial p(w_{ij},z)}{\partial z}\ln\frac{1-p(w_{ij},z)}{p(w_{ij},z)}\nonumber\\
&=&K\sum_{i,j}\frac{w_{ij}}{(1+zw_{ij})^2}\ln\frac{1}{zw_{ij}}
\label{eq:deriv}
\end{eqnarray}
and the second derivative as
\begin{equation}
S''(z)=
K\sum_{i,j}
\frac{w_{ij}}{(1+zw_{ij})^2}\left[-\frac{2w_{ij}}{1+zw_{ij}}\ln\frac{1}{zw_{ij}}-
\frac{1}{z}\right]
\end{equation}
Now let $w_{max}$ denote the maximum weight and $w_{min}$ the minimum \emph{non-zero} weight in the original network.
As $z$ increases from $0$ to $+\infty$, we find that there are five regimes, listed below.

\subsection{$z=0$}
This gives a deterministic ensemble with $p(w_{ij},0)=0$ $\forall i,j$. Therefore the entropy has the minimum value $S(0)=0$, and we are sure that this is not the maximum we are looking for.

\subsection{$0<z< 1/w_{max}$}
Consider first the case $z\ll 1/w_{max}$. 
In this regime, $zw_{ij}\ll 1$ $\forall i,j$, therefore $p(w_{ij},z)\approx zw_{ij}$. So $S(z)$ increases as $z$ increases and no (local) maxima or minima are encountered.
Also in the less strict situation $z< 1/w_{max}$, we have $z< 1/w_{ij}$ $\forall i,j$ which implies $\ln(1/z w_{ij})>0$ $\forall i,j$.
Looking at eq.(\ref{eq:deriv}), this means that $S'(z)>0$, so $S(z)$ increases in the entire range $0<z< 1/w_{max}$.

\subsection{$1/w_{max}<z<1/w_{min}$}
This is the nontrivial range. It can be shown that if a maximum of $S(z)$ exists, it must be within this range. 
As we showed above, when $z<1/w_{max}$ we have $S'(z)>0$.
Similarly, below we will show that when $z>1/w_{min}$ one has $S'(z)<0$.
Taken together, these results imply that, since $S'(z)$ is a continuous function, there must exist a value $z^*$ in the range $1/w_{max}<z^*<1/w_{min}$ such that $S'(z^*)=0$.
As we show later, this corresponds to a maximum of the entropy.


\subsection{$z> 1/w_{min}$}
When $z> 1/w_{min}$, we have $z> 1/w_{ij}$ $\forall i,j$ which implies $\ln(1/z w_{ij})<0$ $\forall i,j$.
Looking at eq.(\ref{eq:deriv}), this means that $S'(z)<0$, so $S(z)$ decreases in the entire range $z> 1/w_{min}$.
Note that in the extreme case $z\gg  1/w_{min}$ we have $zw_{ij}\gg 1$ $\forall i,j$ and $p(w_{ij},z)\approx 1-1/zw_{ij}$. 

\subsection{$z\to+ \infty$}
Now $p(w_{ij},+\infty)=\Theta(w_{ij})$, and the entropy tends again to the minimum value $S(+\infty)=0$.
Interestingly, this limit corresponds to the situation when the original weighted network is regarded as a binary graph by simply setting each non-zero weight to one, and leaving the other values equal to zero.
Within our formalism, we find that this oversimplification corresponds to the minimum entropy, i.e. it is maximally biased.

\section{Real-World Social and Economic Networks\label{sec:real}}
We finally illustrate an application of our method to various real-world social and economic networks.
We consider snapshots of the World Trade Web (WTW), the network of world countries connected by import/export relationships \cite{mypre1,mypre2}, the RyanAir (RA) airport network \cite{RA}, the European Union (EU) aviation network \cite{EU} and the Cond-Mat (CM) scientific collaboration network \cite{CM}.
The WTW is a directed network with no self-loops (hence the number of pairs of vertices is $M=N(N-1)$), the RA and the CM are undirected networks with no self-loops ($M=N(N-1)/2$), and finally the EU is a directed network with self-loops ($M=N(N-1)+N$). 

For each of these networks, we consider the weight matrix $\mathbf{W}$ as given in the original dataset, and use it to calculate the ensemble probabilities defined in eq.(\ref{eq:fermi}) and consequently the entropy $S(z)$ as defined in eq.(\ref{eq:sfinal}).
So the weight $w_{ij}$ is expressed in the (necessarily arbitrary) units used in the original dataset.
We then look for the optimal value $z^*$ that maximizes $S(z)$.
Clearly, $z^*$ corresponds to the optimal weight scale $w^*\equiv 1/z^*$, so that the quantity $z^* w_{ij}$ appearing in eqs.(\ref{eq:fermi}) and (\ref{eq:log}) can be rewritten as 
\begin{equation}
x_{ij}=z ^* w_{ij}=\frac{w_{ij}}{w^*}
\end{equation}
The above expression gives us the optimally rescaled weights $x_{ij}$ of the network, i.e. the weights expressed in terms of the non-arbitrary unit $w^*$.
Note that the rescaled weights $x_{ij}$ are independent of the units used in the data, and hence of the original scale of $w_{ij}$.

\begin{figure}[t]
\centering
\includegraphics[height=6.2cm]{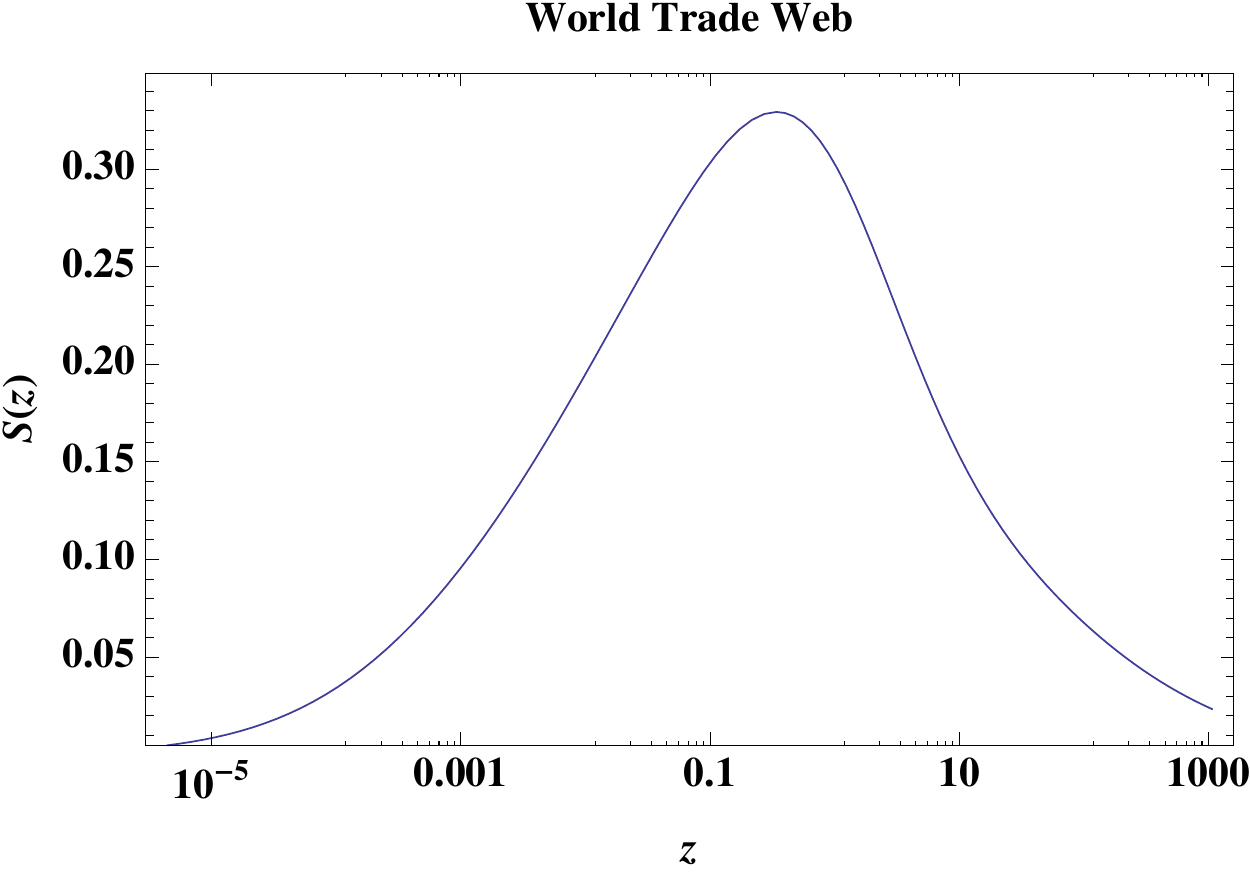}
\caption{Log-linear plot of the ensemble entropy $S(z)$ of the World Trade Web (year 2000) as a function of $z$ in the range $1/w_{max}<z<1/w_{min}$. The number of nodes is 187. The maximum is placed at $z^*_{WTW}=0.34$.}
\label{fig:WTW}
\end{figure}

\begin{figure}[t]
\centering
\includegraphics[height=6.2cm]{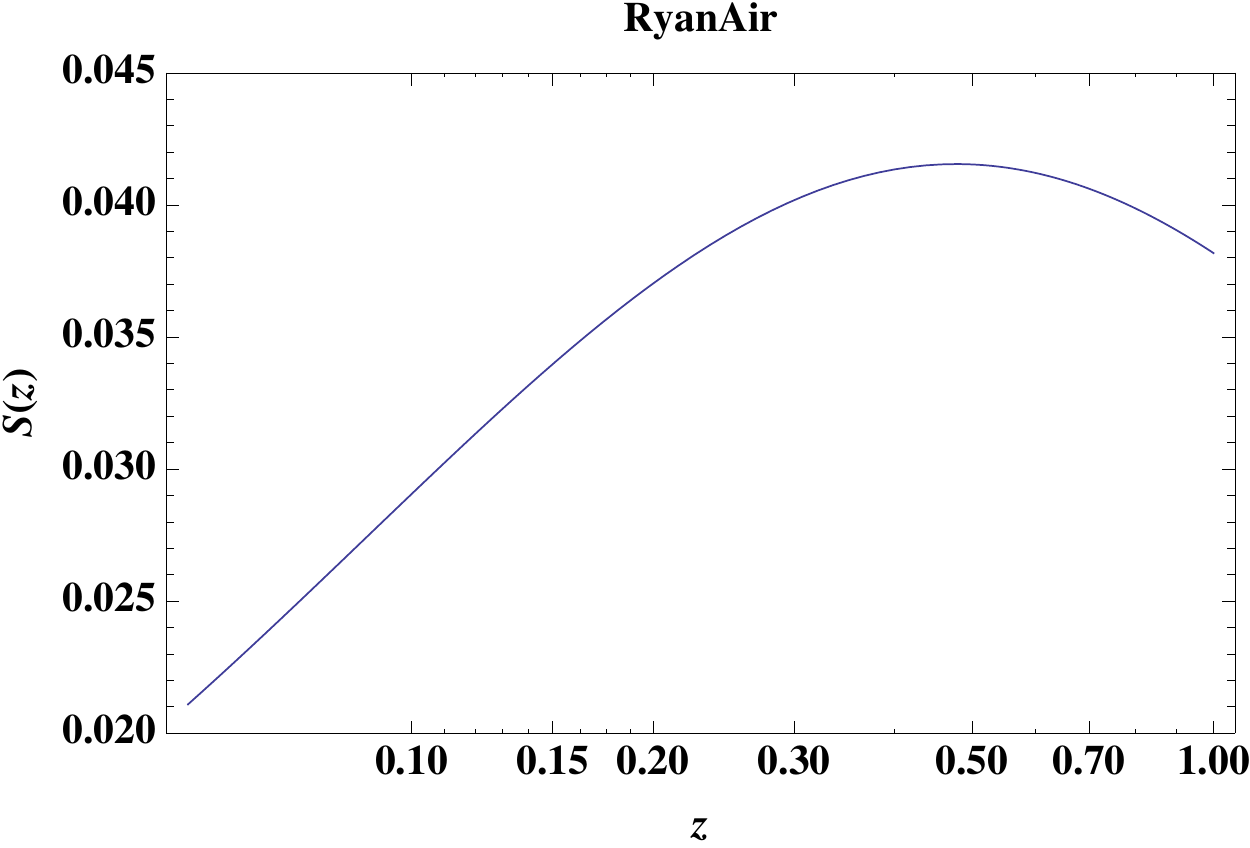}
\caption{Log-linear plot of the ensemble entropy $S(z)$ of the RyanAir network (year 2005) as a function of $z$ in the range $1/w_{max}<z<1/w_{min}$. The number of nodes is 109. The maximum is placed at $z^*_{RA}=0.47$.}
\label{fig:RA}
\end{figure}
\begin{figure}[t]
\centering
\includegraphics[height=6.2cm]{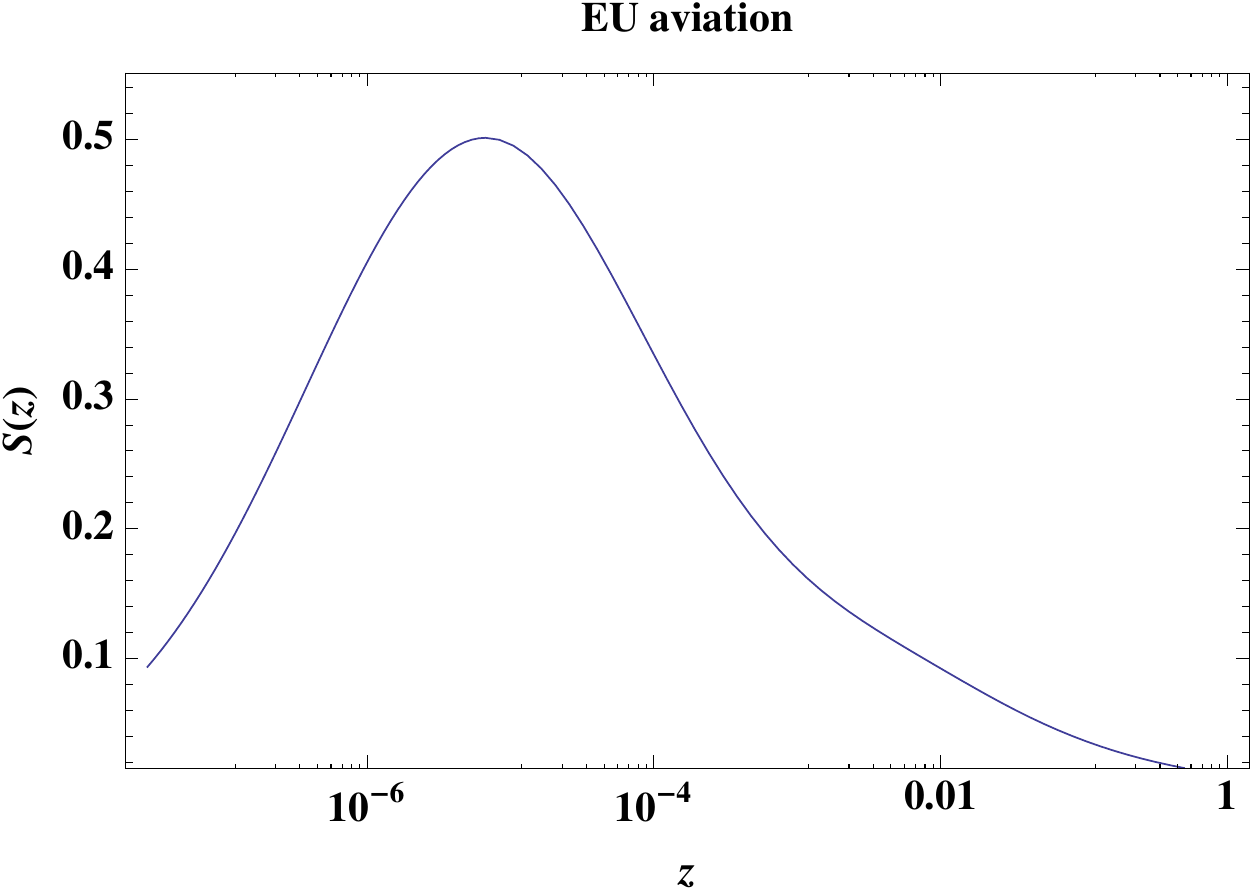}
\caption{Log-linear plot of the ensemble entropy $S(z)$ of the EU aviation network (year 2005) as a function of $z$ in the range $1/w_{max}<z<1/w_{min}$. The number of nodes is 28. The maximum is placed at $z^*_{EU}=6.69\cdot 10^{-6}$.}
\label{fig:EU}
\end{figure}
\begin{figure}[t]
\centering
\includegraphics[height=6.2cm]{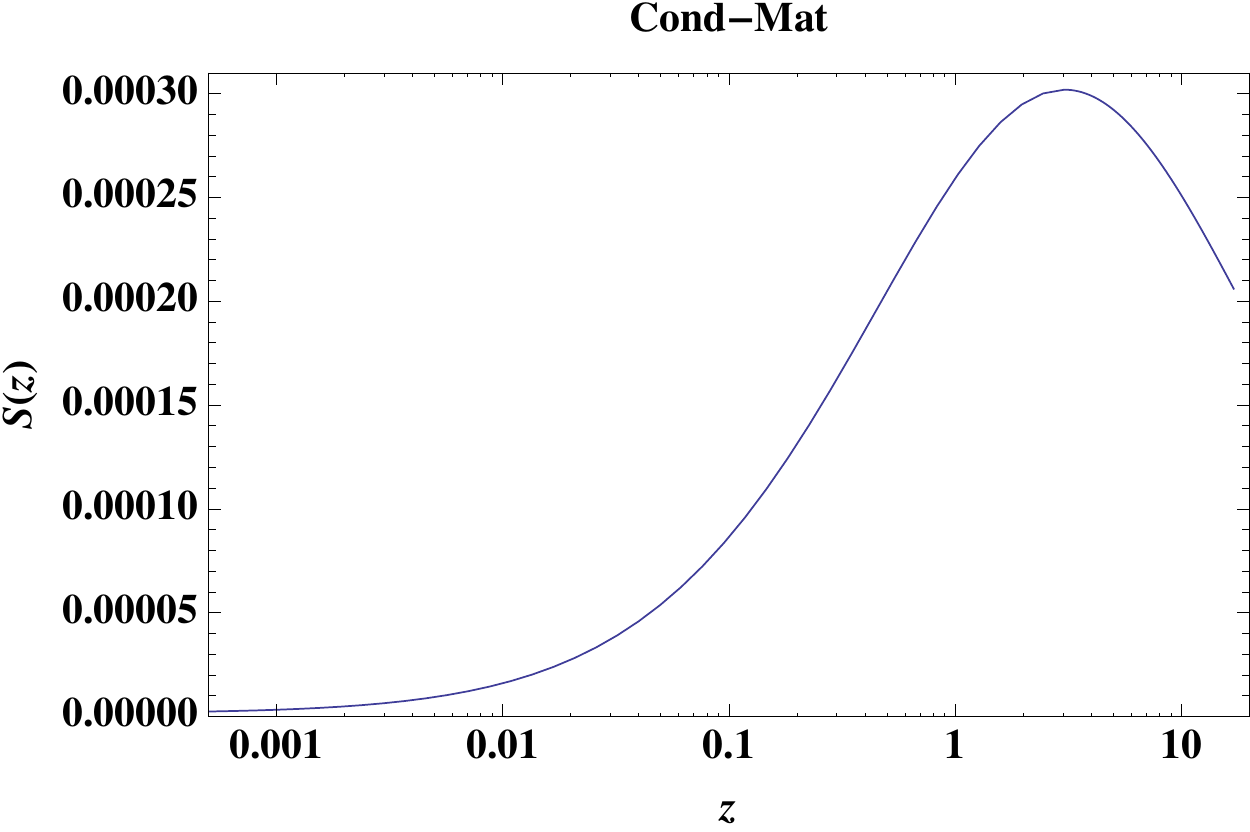}
\caption{Log-linear plot of the ensemble entropy $S(z)$ of the Cond-Mat collaboration network (year 2001) as a function of $z$ in the range $1/w_{max}<z<1/w_{min}$. The number of nodes is 16726. The maximum is placed at $z^*_{CM}=3.034$.} \label{fig:CM}
\end{figure}

The curves of $S(z)$, plotted in the nontrivial range $1/w_{max}<z<1/w_{min}$ where the entropy has a maximum, are shown in fig.\ref{fig:WTW} for the WTW, in fig.\ref{fig:RA} for the RA network, in fig.\ref{fig:EU} for the EU network, and in fig.\ref{fig:CM} for the CM dataset. 
As expected, all curves displays a clear maximum for the value $z^*$ such that $S'(z^*)=0$.
The values of $z^*$ are:
\begin{eqnarray}
z^*_{WTW}&=&0.34\nonumber\\
z^*_{RA}&=&0.47\nonumber\\
z^*_{EU}&=&6.69\cdot 10^{-6}\nonumber\\
z^*_{CM}&=&3.034\nonumber
\end{eqnarray}
The above values give the following optimal units $w^*$ required in order to rescale the original arbitrary matrix $\mathbf{W}$ for each network:
\begin{eqnarray}
w^*_{WTW}&=&2.92\nonumber\\
w^*_{RA}&=&2.09\nonumber\\
w^*_{EU}&=&149355\nonumber\\
w^*_{CM}&=&0.33\nonumber
\end{eqnarray}

Although an analysis of the topological properties of the four networks considered is beyond the scope of this paper, we briefly note that our procedure yields a unique final weight matrix $\mathbf{X}$ expressed in non-arbitrary units, and a corresponding probability matrix $\mathbf{P}(\mathbf{X})$ with entries given by 
\begin{equation}
p_{ij}=p(w_{ij},z^*)=p(x_{ij},1)=\frac{x_{ij}}{1+x_{ij}}\: .
\end{equation}
Using eq.(\ref{eq:mapping}) or (\ref{eq:mapping2}), $\mathbf{P}(\mathbf{X})$ can be finally used in order to compute the least biased weighted generalization $f^{(w)}(\mathbf{X})$ of any binary property $f^{(b)}(\mathbf{A})$.
For instance, for polynomial or multilinear properties 
\begin{equation}
f^{(w)}(\mathbf{X})=f^{(b)}[\mathbf{P}(\mathbf{X})]\: .
\label{eq:super}
\end{equation}
The above formula can be used to compute the otherwise problematic weighted counterparts of many topological properties, e.g. the weighted density and the weighted clustering coefficient mentioned in the Introduction.
For instance, let us consider the ordinary definition of the density $d^{(b)}(\mathbf{A})$ of a binary network $\mathbf{A}$:
\begin{equation}
d^{(b)}(\mathbf{A})\equiv \frac{L(\mathbf{A})}{M}=M^{-1}\sum_{i,j}a_{ij}
\label{eq:density}
\end{equation}
where $L(\mathbf{A})=\sum_{i,j}a_{ij}$ is the total number of links in $\mathbf{A}$ (our usual conventions for $i,j$ in the sum and for the number $M$ of pairs of nodes hold).
Using eq.(\ref{eq:super}), the weighted density of a network with (optimally rescaled) weights $\mathbf{X}$ can be defined as
\begin{equation}
d^{(w)}(\mathbf{X})=d^{(b)}[\mathbf{P}(\mathbf{X})]=M^{-1}\sum_{i,j}p_{ij}=M^{-1}\sum_{i,j}\frac{x_{ij}}{1+x_{ij}} .
\label{eq:wdensity}
\end{equation}
By construction, the above definition takes values between $0$ and $1$, as any proper density measure. This desirable property nicely overcomes the limitations of other naive generalizations of the binary density, illustrating the usefulness of the above approach.

For the four networks in our analysis, the values of the weighted density are:
\begin{eqnarray}
d^{(w)}_{WTW}&=&0.31\nonumber\\
d^{(w)}_{RA}&=&0.022\nonumber\\
d^{(w)}_{EU}&=&0.39\nonumber\\
d^{(w)}_{CM}&=&1.75\cdot 10^{-4}\nonumber
\end{eqnarray}
We stress again that the above values are independent of any (necessarily arbitrary) choice of the unit of weight in the orginal data. 
It is interesting to compare the above values of the weighted density $d^{(w)}$ with the corresponding values of the ordinary binary density $d^{(b)}$, as measured on the adjacency matrix characterizing the bare topology of the original network:
\begin{eqnarray}
d^{(b)}_{WTW}&=&0.58\nonumber\\
d^{(b)}_{RA}&=&0.044\nonumber\\
d^{(b)}_{EU}&=&0.86\nonumber\\
d^{(b)}_{CM}&=&3.42\cdot 10^{-4}\nonumber
\end{eqnarray}
We find that the values of $d^{(b)}$ for all networks are approximately twice the corresponding values of $d^{(w)}$.
This big numerical difference shows the entity of the information loss encountered when a weighted network is regarded as a binary one (corresponding to the maximally biased limit $z\to\infty$ as discussed in sec. \ref{sec:maxent}). 
Our approach instead makes use of all the available information encapsulated in the weights, and ensures that the bias is minimized (corresponding to the maximum-entropy point $z^*$).
For the four networks in our analysis,  exploiting the additional knowledge of the weights has a significant `sparsifying' effect, approximately halving the purely binary density.

\section{Conclusions}
In this paper we have addressed two problems of abitrariness that are systematically encountered in the analysis of weighted networks: the non-uniqueness of the generalization of binary topological properties to their weighted counterparts and that of the scale of edge weights.
While in principle independent, we have shown that, when a weighted network is mapped to an ensemble of binary graphs, these two problems turn out to be intimately related.
In particular, the ensemble formalism (especially when rewritten within a statistical-physics framework) provides a straightforward weighted generalization of any binary property, and at the same time allows us to find the optimal weight scale via a Maximum Entropy criterion.
It is remarkable that such a criterion cannot be invoked directly on the original system by maximizing the entropy of the weighted network, because the entropy is only defined for ensembles of graphs and not for a single instance (unless the original weighted network is trivially regarded as the only possible outcome of a deterministic ensemble with zero entropy).
Therefore the transition from a single network to an ensemble of graphs is necessary in order to find the least biased scale of weights via a maximization of the entropy.
Using examples of real-world socio-economic networks, we have illustrated our approach and computed the optimal scale for such networks.
We have shown that this scale can be used to define the least biased generalization of any binary property to the weighted case, confirming that the problem of selecting an optimal scale and that of defining unique generalizations of binary properties are tightly interrelated within our ensemble formalism.

\subsubsection*{Acknowledgments.} D.G. acknowledges support from MULTIPLEX (317532) and the Dutch Econophysics Foundation (Stichting Econophysics, Leiden, the Netherlands) with funds from beneficiaries of Duyfken Trading Knowledge BV, Amsterdam, the Netherlands.
S.E.A. was supported by The Leverhulme Trust, UK and The Royal Society, UK.
G.C. acknowledges support from FET project FOC (255987) and MULTIPLEX (317532).

\end{document}